# Doubly discrete Lagrangian systems related to the Hirota and Sine-Gordon equation

Claudio Emmrich[*] and Nadja Kutz[**]
Technische Universität Berlin
Sekr. MA 8-5, Str. des 17. Juni 136
10623 Berlin, Germany

21.9.1994

## Abstract

We extend the action for evolution equations of KdV and MKdV type which was derived in [NC] to the case of not periodic, but only equivariant phase space variables, introduced in [FV2]. The difference of these variables may be interpreted as reduced phase space variables via a Marsden-Weinstein reduction where the monodromies play the role of the momentum map. As an example we obtain the doubly discrete sine-Gordon equation and the Hirota equation and the corresponding symplectic structures.

## 1 Introduction

We will investigate a class of integrable lattice systems (see e.g. [BRST, NCP, S, V]) defined by an evolution equation of the following type:

$$u_u - u_d - V'(u_l - u_r) = 0, \qquad (1)$$

where $u$, $d$, $l$ and $r$ denote up, down, left and right, respectively and $V'(x)$ is the derivative of $V(x) : \mathbb{R} \to \mathbb{R}$. If we start with initial data on a Cauchy path $\mathcal{C}$ on a light cone lattice the local evolution given by (1) will determine the function $u$ on the whole lattice.

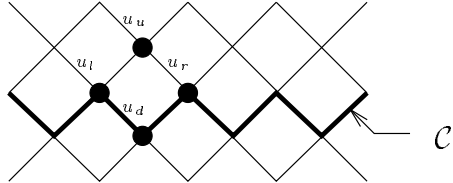

An important application of the system above appeared in the study of doubly discrete variants of the Sine-Gordon equation. Here one obtains the well known Hirota equation [H], which is equivalent to (1) for $V'(x) = -i \ln(\frac{1+ke^{ix}}{k+e^{ix}})$ if one redefines $u \to -u$ along every second diagonal of the light cone lattice (see [FV1, FV2]). The difference of two adjacent values of $u$, i.e. $x_{t,k} := u_{t,k-1} - u_{t,k+1}$ (see fig.1) for the potential above describes (modulo a redefinition along the diagonals [BKP])

---

Supported by the Deutsche Forschungsgemeinschaft, Sonderforschungsbereich 288
[*]e-mail: cemm@phyq2.physik.uni-freiburg.de
[**]e-mail: nadja@math.tu-berlin.de

the angle between the tangent directions of a discrete K-surface in the Tchebycheff parametrization [BP]. Both equations possess quantum analogs ([BKP],[FV2]), where the equation of motion for the "quantized" angles of a K-surface proves to be a beautiful example in the theory of quantum groups [BBR].

In the following we show that it is possible to derive both types of equations as equation of motions from one action, which will be an extended version of the one given in [NC]. The main new feature of this action is that we admit non-vanishing monodromies which become dynamical variables. This extended action leads to equations of motion which exactly coincide with the evolution equations (1), whereas the action in [NC] only leads to the *difference* of the evolution equations for adjacent faces, and not to the evolution equation on a single face.

An explanation of the above features in terms of symplectic geometry will be given in the last section.

## 2 The phase space

A (spatially) periodic light cone lattice $L_{2p}$ with period $2p$ may be viewed as $L/\mathbb{Z}$, where $\mathbb{Z}$ acts on the infinite light cone lattice $L$ by shifts by $2p$ in space-like direction (cf. fig. 1). A quasi-periodic field is a mapping

$$u : L \to \mathbb{R}$$

with

$$u_{t,i+2p} - u_{t,i} = u_{t,i+2p+2} - u_{t,i+2} \quad \forall i,$$

i.e., there are two (space independent) monodromies $C_t^{(1)}, C_t^{(2)}$ defined by $C_t^{(i)} = u_{t,2p+2k+1-i} - u_{t,2k+1-i}$ for an arbitrary $k \in \mathbb{Z}$.

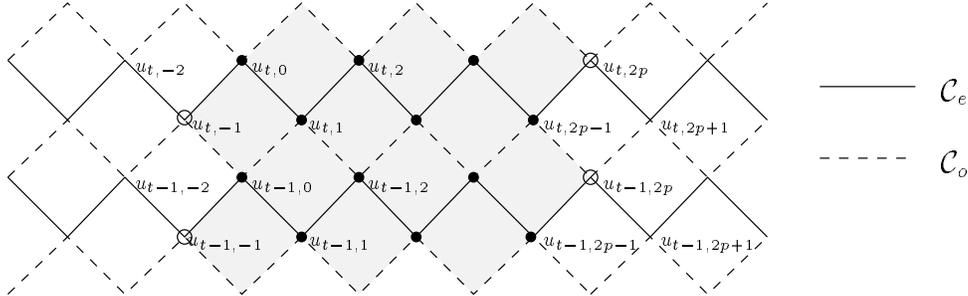

Figure 1

We denote by $\mathcal{P}$ the set of faces on a fundamental domain of the lattice (cf. fig. 1). On the set of quasi-periodic fields we define the following action:

$$\begin{aligned}
S &= \sum_{P \in \mathcal{P}} \{u_d(P)(u_l(P) - u_r(P)) + V(u_l(P) - u_r(P))\} \\
&\quad + \sum_t \left\{ u_{t+1,2p-1} C_t^{(1)} - \frac{1}{2} C_t^{(1)} C_{t+1}^{(2)} + u_{t,0} C_t^{(2)} + \frac{1}{2} C_t^{(1)} C_t^{(2)} \right\} \qquad (2) \\
&= \sum_t \Bigg\{ \sum_{k=1}^{p-1} u_{t,2k}(u_{t+1,2k-1} - u_{t+1,2k+1}) + u_{t,0}(u_{t+1,2p-1} - C_{t+1}^{(2)} - u_{t+1,1}) \\
&\quad + u_{t+1,2p-1} C_t^{(1)} - \frac{1}{2} C_t^{(1)} C_{t+1}^{(2)} \\
&\quad + \sum_{k=0}^{p-2} u_{t,2k+1}(u_{t,2k} - u_{t+1,2k+2}) + u_{t,2p-1}(u_{t,2p-2} - u_{t,0} - C_t^{(1)}) \Bigg\} \qquad (3)
\end{aligned}$$

$$+u_{t,0}C_t^{(2)} + \frac{1}{2}C_t^{(1)}C_t^{(2)} + \sum_{k=0}^{2p-3} V(u_{t,k} - u_{t,k+2})$$
$$+V(u_{t,2p-1} - C_t^{(2)} - u_{t,1}) + V(u_{t,2p-2} - u_{t,0} - C_t^{(1)})\Big\}$$

This action differs from the action in [NC], which is only defined on periodic fields, by the addition of the extra terms in the second line of equation (2). Those extra terms guarantee that the action is invariant under a space-like shift, and hence is independent of the choice of the fundamental domain. Note, that the monodromies entering the action may a priori be time dependent: They become dynamical variables rather than pure parameters, and their time independency will be a consequence of the equations of motion.

$\{(u_{t,k})_{t\in\mathbb{Z}, k\in\{-1,\ldots 2p\}}\}$ and $\{(u_{t,k})_{t\in\mathbb{Z}, k\in 0,\ldots 2p-1}, C_t^{(1)}, C_t^{(2)}\}$ form independent coordinate systems on the set of quasi-periodic fields. Whereas the former is particularly useful for the derivation of the symplectic structure in the next section, the equations of motion may be most easily derived using the latter coordinate system.

Variation with respect to $u_{t,k}, (t\in\mathbb{Z}, k\in 0,\ldots 2p-1)$, yields (exactly as in [NC]) the difference of the evolution equations (1) for the neighbouring faces to the left and to the right of $u_{t,k}$:

$k$ even:
$$u_{t+1,k-1} - u_{t,k-1} - V'(u_{t,k-2} - u_{t,k}) - u_{t+1,k+1} - u_{t,k+1} - V'(u_{t,k} - u_{t,k+2}) = 0,$$
$k$ odd:
$$u_{t,k-1} - u_{t-1,k-1} - V'(u_{t,k-2} - u_{t,k}) - u_{t,k+1} - u_{t-1,k+1} - V'(u_{t,k} - u_{t,k+2}) = 0$$
(4)

Variation with respect to the monodromies $C_t^{(i)}$, $i = 1, 2$ yields:

$$\begin{aligned} u_{t+1,2p-1} - u_{t,2p-1} - V'(u_{t,2p-2} - u_{t,0} - C_t^{(1)}) &= \tfrac{1}{2}C_{t+1}^{(2)} - \tfrac{1}{2}C_t^{(2)} \\ u_{t,0} - u_{t-1,0} - V'(u_{t,2p-1} - C_t^{(2)} - u_{t,1}) &= \tfrac{1}{2}C_{t-1}^{(1)} - \tfrac{1}{2}C_t^{(1)}. \end{aligned}$$
(5)

Now, as we may write the monodromy as the sum of differences of field variables:

$$C_t^{(i)} = \sum_{k=1}^p (u_{t,2k+1-i} - u_{t,2k-1-i})$$

equations (4) are sufficient to enforce the time independence of the monodromies. Hence, the right hand sides of equations (5) vanish. We get the evolution equations (1) for the faces "above" $u_{t,2p-1}$ and $u_{t,0}$ for all $t$, and thus finally for all faces.

## 3 Symplectic Structure

Using straightforward modifications of covariant phase space techniques [Z, CW], we first show how one can derive a symplectic structure from any action on a light cone lattice which may be written as the sum of terms whith support on the "canonical Cauchy paths" $\mathcal{C}_{o,e}^t$ (cf. fig. 2). This symplectic structure will automatically be invariant under evolution.

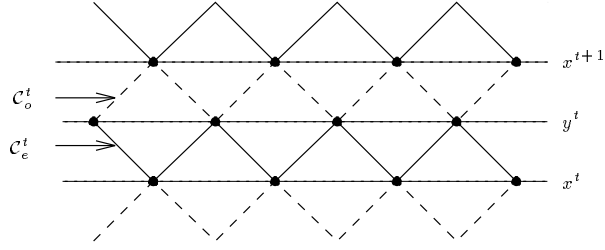

Figure 2

There is in general no canonical way of identifying $\mathcal{C}_o^t$ with $\mathcal{C}_e^t$. Therefore, our ansatz for the action is:

$$S = \sum_t \left( L_1(\mathbf{x}^t, \mathbf{y}^t) + L_2(\mathbf{y}^t, \mathbf{x}^{t+1}) \right)$$

with two different functions $L_1$ and $L_2$. Here, $\mathbf{x}^t$ are the variables on the lower vertices of $\mathcal{C}_e^t$, and $\mathbf{y}^t$ the variables on the upper vertices of $\mathcal{C}_e^t$ (or, equivalently, the lower vertices of $\mathcal{C}_o^t$).

Variation of the action with respect to $\mathbf{x}^t$ and $\mathbf{y}^t$ yields the equations of motion:

$$\begin{aligned} B_1^t &:= D_1 L_1(\mathbf{x}^t, \mathbf{y}^t) + D_2 L_2(\mathbf{y}^{t-1}, \mathbf{x}^t) = 0 \\ B_2^t &:= D_2 L_1(\mathbf{x}^t, \mathbf{y}^t) + D_1 L_2(\mathbf{y}^t, \mathbf{x}^{t+1}) = 0, \end{aligned}$$

where $D_{1,2}$ denote the derivatives with respect to the first and second argument, respectively.

We may consider $L_1(\mathbf{x}^t, \mathbf{y}^t)$ as a functional on covariant phase space, i.e., the space of all solutions to the equations of motion. Denoting by $\mathbf{d}$ the exterior derivative on covariant phase space, we get:

$$\begin{aligned} \mathbf{d} L_1(\mathbf{x}^t, \mathbf{y}^t) &= -\mathbf{d} L_2(\mathbf{y}^{t-1}, \mathbf{x}^t) + B_1^t \mathbf{dx}^t + B_2^t \mathbf{dy}^{t-1} \\ &\quad - D_2 L_1(\mathbf{x}^{t-1}, \mathbf{y}^{t-1}) \mathbf{dy}^{t-1} + D_2 L_1(\mathbf{x}^t, \mathbf{y}^t) \mathbf{dy}^t \end{aligned} \quad (6)$$

On covariant phase space $B_{1,2}^t$ vanish. Hence:

$$0 = \mathbf{d}^2 L_1(\mathbf{x}^t, \mathbf{y}^t) = \mathbf{d}\left( D_2 L_1(\mathbf{x}^t, \mathbf{y}^t) \mathbf{dy}^t \right) - \mathbf{d}\left( D_2 L_1(\mathbf{x}^{t-1}, \mathbf{y}^{t-1}) \mathbf{dy}^{t-1} \right)$$

Thus, $\omega := -\frac{1}{2}\mathbf{d}\left( D_2 L_1(\mathbf{x}^t, \mathbf{y}^t) \mathbf{dy}^t \right)$ defines a presymplectic structure (i.e., a closed, possibly degenerate two-form on phase space) which is invariant under the evolution. This presymplectic structure will be symplectic if there is no gauge symmetry [BHS]. Similarly, we may define another symplectic structure using $L_2$.

If, as in the applications we have in mind, the functions $L_i$ are invariant under translations in space directions, it makes sense to compare $L_1$ and $L_2$, by identifying $\mathcal{C}_e^t$ with $\mathcal{C}_o^t$ by half-shift in light-cone direction either to the right or to the left: Due to the translation invariance, both identifications yield the same result.

For the action (2) we may choose $L_1 = L_2 =: L$ with

$$L(\mathbf{v}, \mathbf{w}) = \sum_{k=0}^{p-1} v_k(w_k - w_{k+1}) + V(w_k - w_{k+1}) + w_0(v_{p-1} - v_{-1}) + \frac{1}{2}(v_{p-1} - v_{-1})(w_p - w_0)$$

and the identifications

$$v_k = u_{t,2k+1}, \qquad w_k = u_{t,2k}$$

for $L_1$, and
$$v_k = u_{t,2k}, \qquad w_k = u_{t+1,2k-1}$$
for $L_2$.

Thus, we get a translation invariant symplectic structure

$$\begin{aligned}\omega = &-\frac{1}{2}\Big(\sum_{k=0}^{p-1} \mathrm{d}u_{t,2k+1} \wedge (\mathrm{d}u_{t,2k} - \mathrm{d}u_{t,2k+2}) \\ &+ (\mathrm{d}u_{t,2p-1} - \mathrm{d}u_{t,-1}) \wedge \mathrm{d}u_{t,0} + \frac{1}{2}(\mathrm{d}u_{t,2p-1} - \mathrm{d}u_{t,-1}) \wedge (\mathrm{d}u_{t,2p} - \mathrm{d}u_{t,0})\Big) \\ = &-\frac{1}{2}\Big(\sum_{k=0}^{p-2} \mathrm{d}u_{t,2k+1} \wedge (\mathrm{d}u_{t,2k} - \mathrm{d}u_{t,2k+2}) + \mathrm{d}u_{t,2p-1} \wedge (\mathrm{d}u_{t,2p-2} - \mathrm{d}u_{t,0} - \mathrm{d}C_t^{(1)}) \\ & + \mathrm{d}C_t^{(2)} \wedge \mathrm{d}u_{t,0} + \frac{1}{2}\mathrm{d}C_t^{(2)} \wedge \mathrm{d}C_t^{(1)}\Big)\end{aligned}$$

for $\mathcal{C}_e^t$, and similarly for $\mathcal{C}_o^t$. The corresponding Poisson structure exactly coincides with the one given in [FV2]:

$$\begin{aligned}\{u_i^t, u_k^t\} &= 0, & &\text{if } i-k \text{ even} \\ \{u_i^t, u_k^t\} &= 1, & &\text{if } i-k \text{ odd}, i < k, |i-k| < 2p, \\ \{u_i^t, C_t^{(k)}\} &= 0 & &\text{if } i-k \text{ odd}, \\ \{u_i^t, C_t^{(k)}\} &= 2 & &\text{if } i-k \text{ even} \\ \{C_t^{(1)}, C_t^{(2)}\} &= 0 & & \end{aligned} \qquad (7)$$

## 4 Marsden-Weinstein reduction

The process of going over from the variables $u_{t,k}$ to the difference variables $x_{t,k} := u_{t,k-1} - u_{t,k+1}$ may be interpreted as a Marsden Weinstein reduction [AM, GS]: If we identify the covariant phase space with the space $M$ of quasi-periodic initial conditions on a Cauchy path $\mathcal{C}_e^t$, then an action of $\mathbf{G} = \mathbb{R} \times \mathbb{R}$ on $M$ which commutes with time evolution is defined by:

$$\begin{aligned}(\alpha,\beta) \cdot u_{t,2k} &:= u_{t,2k} + \alpha \\ (\alpha,\beta) \cdot u_{t,2k+1} &:= u_{t,2k} + \beta\end{aligned}$$

for arbitrary $k \in \mathbb{Z}$. To this action, there exists a momentum map $\mathcal{J} : M \to \mathcal{G}^* = \mathbb{R} \oplus \mathbb{R}$ whose components are simply

$$\mathcal{J}_{(1,0)} = \frac{1}{2} C_t^{(2)}, \quad \mathcal{J}_{(0,1)} = \frac{1}{2} C_t^{(1)}.$$

This momentum map is trivially $Ad^*$-equivariant, as the group is abelian and $C_t^{(1)}, C_t^{(2)}$ Poisson-commute. Hence, for arbitrary $\rho, \mu \in \mathbb{R}$, we may consider the Marsden-Weinstein reduced phase-space

$$M_{(\rho,\mu)} = \mathcal{J}^{-1}((\rho,\mu))/(\mathbb{R} \times \mathbb{R})$$

which is a symplectic manifold again: It is just "the space of the difference variables $x$ for fixed monodromy".

Alternatively, one may describe the reduced phase space as $\mathcal{J}^{M \times \mathcal{O}^-}(0)/\mathbf{G}$, where $\mathcal{O}^-$ is a coadjoint orbit with minus its standard symplectic structure, and $\mathcal{J}^{M \times \mathcal{O}^-}$ is the momentum map on the enlarged phase space $M \times \mathcal{O}^-$ [GS]. As the group is abelian, the coadjoint orbit in our situation is simply one point. Nevertheless, this picture is useful for understanding the structure of the space $X$ obtained by gluing all the reduced phase spaces $M_{(\rho,\mu)}$ together:

The space obtained in this way is "the space of the difference variables $x$ for arbitrary monodromy". It may be described as $\Phi^{-1}(0)/\mathbf{G}$, where $\Phi^{-1}(0) \subset M \times \mathcal{G}^*$ and $\Phi$ is the continuation of $\mathcal{J}^{M \times \mathcal{O}^-}$ to $M \times \mathcal{G}^*$ (i.e., $\mathcal{J}^{M \times \mathcal{O}^-} = \Phi|_{M \times \mathcal{O}^-}$ for all coadjoint orbits $\mathcal{O} \subset \mathcal{G}^*$).

Here, $M \times \mathcal{G}^*$ is no longer a symplectic, but only a Poisson manifold. The monodromies as functions on $X$ are induced by the (trivially) $Ad^*$-invariant functions $(\rho, \mu) \mapsto \rho$ and $(\rho, \mu) \mapsto \mu$ on $\mathcal{G}^*$.

Although we are only dealing with a rather trivial application of the reduced phase space techniques, this symplectic reformulation of the transition from the variables $u$ to the variables $x$ sheds light on the role of the monodromies as dynamical variables. In particular, it naturally explains why the monodromies have non-trivial Poisson-brackets with the variables $u$, but are in the center of the reduced Poisson algebra [K].

This should be a guideline for applications to non-abelian situations, which for example appear in a directly related way in the evolution equation for the Gauss map for discrete K-surfaces [BP].

# 5 Acknowledgements

We would like to thank U. Pinkall, A. Bobenko and F. Faddeev for many helpful discussions.